\documentclass[twocolumn]{aastex62}
\usepackage{amsmath}
\usepackage{mathtools}

\graphicspath{{figures/}}

\def\mean#1{\left< #1 \right>}
\def\ed#1{{#1}}

\received{}
\revised{}
\accepted{}

\shorttitle{Magnetic grain alignment in PPDs}
\shortauthors{Yang H.}

\begin{document}

\title{Size limit of superparamagnetic inclusions in dust grains and difficulty of magnetic grain alignment in protoplanetary disks}

\correspondingauthor{Haifeng Yang}
\email{yanghaifeng@tsinghua.edu.cn}

\author[0000-0002-8537-6669]{Haifeng Yang}
\altaffiliation{C.N. Yang Junior Fellow}
\affil{Institute for Advanced Study, Tsinghua University, Beijing, 100084, China}




\begin{abstract}
Alignment of non-spherical grains with magnetic fields is an important problem as it lays the foundation of probing magnetic 
fields with polarized dust thermal emissions. In this paper, we investigate the feasibility of magnetic
alignment in protoplanetary disks (PPDs). We use an alignment condition
that Larmor precession should be fast compared with the damping timescale. We first show that 
the Larmor precession timescale is some three orders of magnitude longer than the damping time for millimeter-sized grains 
under conditions typical of PPDs, making the magnetic alignment unlikely. The precession time can be shortened by 
superparamagnetic inclusions (SPIs), but the reduction factor strongly depends on the size of the SPI clusters, which we 
find is limited by the so-called ``N\'{e}el's relaxation process." In particular, the size limit of SPIs is set by the so-called 
``anisotropic energy constant" of the SPI material, which describes the energy barrier needed to change the direction of the 
magnetic moment of an SPI. For the most common iron-bearing materials, we find maximum SPI sizes 
corresponding to a reduction factor of the Larmor precession timescale of order $10^3$. 
We also find that reaching this maximum reduction factor requires fine-tuning \ed{on the SPI sizes}.
Lastly, we illustrate the effects of the SPI size limits on 
magnetic alignment of dust grains with a simple disk model, and we conclude that it is unlikely for relatively large grains of 
order 100 $\mu$m or more to be aligned with magnetic fields even with SPIs.

\end{abstract}



\section{Introduction} \label{sec:intro}

The polarization of starlight was first observed in 1949 \citep{Hiltner1949nature}. It was soon 
attributed to the alignment of dust grains in the foreground interstellar medium. Since then, the
alignment of dust grains, especially with respect to magnetic fields, has many
developments. 
Many theories were developed to explain how dust grains are aligned with magnetic fields, such as
Davis-Greenstein mechanism \citep{DG1951}, hydrogen formation torque \citep{Purcell1979}, radiative 
alignment torque (B-RAT; \citealt{DM1976,DW1997,LH2007}), and recently mechanical alignment torque 
(B-MAT; \citealt{Hoang2018}). 
We refer interested readers to \cite{Andersson2015} \ed{and references therein}. 

Superparamagnetism (SPM) was first introduced to the astronomical literatures of grain alignment by \cite{JS1967}.
They pointed out that SPM can enhance magnetic relaxation, the process invoked by Davis-Greenstein mechanism to dissipate
oscillating magnetic moments and to align grains, which was found 
insufficient \ed{to align} regular paramagnetic dust grains \ed{with magnetic fields}. 
\cite{Mathis1986} adopted this theory with the assumption that grains containing any small superparamagnetic particle, the so-called 
superparamagnetic inclusions (SPIs),
can be aligned with magnetic fields. Under this theory, the fact that bigger grains are better aligned
is well explained since bigger grains are more likely to contain SPIs.
Fe-Ni inclusions appear to present in interplanetary dust particles, and their spatial frequency supports Mathis's 
theory in explaining the wavelength dependence of polarization \citep{Goodman1995}.
Magnetic nanoparticles and inclusions were also discussed recently by \cite{DraineHensley2013} focusing on the 
impacts of such inclusions on the dust thermal emission and polarization.

Observationally, tracing magnetic fields with polarized
thermal emission \ed{from grains aligned with magnetic fields} is a classical and 
successful method. 
It is clear that grains in diffuse interstellar medium are aligned with magnetic field 
\ed{from starlight polarization} \citep{Mathewson1970}. 
Recently, this picture receives firm supports from the Planck all-sky 
survey data \citep{Planck2015XIX}, and from the interferomatric polarimetry data (see \citealt{Hull2019} and 
reference therein). It is in no doubt that grains on scales larger than disks are aligned with magnetic fields.

In the past six years, thanks to the improvement of interferomatric polarimetry, especially with the Atacama Large Millimeter/submillimeter Array (ALMA), we have become able to resolve
polarization maps down to the disk scale. The results have been surprising. Most systems show
uniform polarization patterns, especially at shorter wavelengths, e.g. HL Tau \citep{Stephens2014,Stephens2017}, IM Lup \citep{Hull2018}, DG Tau \citep{Bacciotti2018}, and HD163296 \citep{Dent2019}, which are
better explained with the self-scattering of dust grains than with magnetically aligned grains
\citep{Kataoka2015,Yang2016a,Ohashi2019,Lin2019}. \cite{Cox2018}'s survey found a trend that the polarization is uniform on scales smaller 
than 100 AU at percent level, whereas the polarization is less organized on larger scales at a higher 
level ($\sim 5\%$). This is consistent with the picture that dust grains at disk scales are not aligned. 
There are also some systems showing complicated polarization features, such as the binary system BHB07-11 \citep{Alves2018}, the southern part of HD142527 \citep{Kataoka2016b,Ohashi2018}, and HL Tau \citep{Kataoka2017}.
However, the origin of these additional complexities is not certain at this time.

Theoretically, the transition from magnetic alignment to no-magnetic-alignment going from the more diffuse surrounding region to the disk is not surprising. 
The protoplanetary disk (PPD) environment is very different from 
the diffuse ISM in many ways: higher density, bigger grain sizes, different temperature and radiation
energy density, etc. A rough comparison given in Sec.~\ref{sec:basics} shows that the Larmor precession
timescale can be some three orders of magnitude larger than the gaseous damping timescale, making magnetic alignment impossible
(see also \citealt{Tazaki2017}).
\cite{Hoang2017} pointed out that if the dust grains contain large SPIs (on the order of $10^5$ iron atoms each), magnetic alignment of millimeter dust grains can still be
possible in PPDs. The enhancement from SPIs strongly depends on the size of each SPI. In this paper
we will focus on estimating the maximum size of SPIs and discuss how the size limit affects the magnetic 
alignment of dust grains in PPDs. 

The structure of this paper is as follows. In Sec.~\ref{sec:basics}, we introduce the basic 
timescales involved in the magnetic alignment process, and give a simple comparison for regular paramagnetic
materials. In Sec.~\ref{sec:spi}, we introduce how SPIs can enhance the magnetic susceptibility and how 
the cluster size is limited by the relaxation process.
In Sec.~\ref{sec:spisize}, we estimate the 
size limit of SPIs assuming a few commonly adopted forms of iron. 
In Sec.~\ref{sec:disk}, we adopt a simple disk
model and discuss the \ed{feasibility} of magnetic alignment in \ed{PPDs} In Sec.~\ref{sec:discussion}, 
we discuss caveats and uncertainties in this work, as well as its implications for several observed disks. 
\ed{We summarize our results} in Sec.~\ref{sec:summary}.

\section{Basic magnetic alignment theory}
\label{sec:basics}
\subsection{Criteria for magnetic alignment}
\label{subsec:criteria}
Dust grain alignment with magnetic fields is a complicated process with many timescales involved. An outline of the process
(regardless of alignment mechanism) is as follows (see e.g. \citealt{Lazarian2007}).
Initially, a randomly-spinning dust grain nutates about its angular momentum axis. Over a timescale $t_\mathrm{int}$, the
principal axis of the dust grain becomes aligned with the its angular momentum through some relaxation processes. 
The dust grain's angular momentum also precesses around the external magnetic field on the 
Larmor precession timescale ($t_L$). 
At the end, some torques gradually force the angular momentum of the dust grain to be aligned 
with the magnetic field, over a timescale of $t_\mathrm{al}$.
At the same time, the random bombardment of gas particles tries to disturb the angular momentum, over the gaseous
damping timescale ($t_{d}$).

\ed{One typically considers that these three timescales must follow a hierarchical in equality
-- i.e., $t_L<t_\mathrm{al}<t_d$ -- in order for grain alignment to proceed successfully. 
However, the alignment timescale $t_\mathrm{al}$ is considerably more complicated than the
other two timescales to compute in general, because it depends on the specific torquing 
mechanism at play. For the purpose of this paper, we work with the insufficient but
still necessary condition $t_L<t_d$ to determine conditions under which dust grains can
align with ambient magnetic fields}\footnote{Note that this criterion effectively ignored the suprathermal rotation, which 
will be discussed in Sec.~\ref{sec:discussion}.}. 
Even though \ed{$t_L >t_d$ is unrealistic} in most interstellar medium, it becomes 
a strong possibility in PPDs as grains grow to millimeter in size and gas density increases 
by many orders of magnitude compared with the ISM values, as we will see more quantitatively 
in the following section.

In theoretical works studying the dynamics of magnetic grain alignment, fast Larmor precession
has been mostly assumed, so that the calculated torques are averaged over one precession before 
the dynamics of the dust grain is studied (e.g., \citealt{LH2007,Hoang2018}). 
In this work, we use $t_L<0.1t_d$ as the criterion for fast Larmor precession 
assumption. In this regime, the Larmor precession timescale will not be the limiting
factor, and magnetic grain alignment becomes possible. We will come back to discuss this criterion
and its caveats in more detail in Sec.~\ref{sec:discussion}.

\subsection{Gaseous damping timescale}
The random bombardment of gas particles on dust grains tends to misalign any ordered orientation 
of the dust grains. 
\ed{This happens roughly on a timescale of \citep{Roberge1993}}:
\begin{equation}
    \begin{split}
    t_\mathrm{d} =& \frac{2\sqrt{\pi}}{5}\frac{\rho_s a}
    {n_{\rm g}m_{\rm g}v_{\rm g,th}}\\
    =& 3.54\times 10^{12}~\mathrm{s}\times \left(\frac{\rho_s}{3~\mathrm{g/cm^3}}\right)\\
    &\times \left(\frac{a}{0.1~\mathrm{\mu m}}\right)\left(\frac{n_{\rm g}}{20~\mathrm{cm^{-3}}}\right)^{-1}\left(\frac{T_g}{85~\mathrm{K}}\right)^{-1/2},
    \end{split}
\end{equation}
where $\rho_s$ is the mass density of the (solid) dust grain, $a$ is the grain size, $n_g$ is the number density of 
gas particles, and $T_g$ is the gas temperature.
This is a rather long time scale for the diffuse ISM conditions we used above. 
If we take a PPD with total mass of $0.01\rm\,M_\sun$ as normalization, 
uniformly distributed in a cylinder with 
$100$ AU as radius and $10$ AU as scale height,
we have:
\begin{equation}
    \label{eq:td}
    \begin{split}
    t_\mathrm{d} 
    =& 2.6\times 10^{8}~\mathrm{s}\times \left(\frac{\rho_s}{3~\mathrm{g/cm^3}}\right)\left(\frac{a}{1~\mathrm{mm}}\right)\\
    &\times \left(\frac{n_{\rm g}}{5\times 10^9~\mathrm{cm^{-3}}}\right)^{-1}\left(\frac{T_g}{25~\mathrm{K}}\right)^{-1/2},
    \end{split}
\end{equation}
\ed{which} is \ed{shorter} than typical dynamical timescales in PPDs.

\subsection{Larmor precession timescale}
The Larmor precession process is the precession of a magnetic moment around a magnetic field. 
A spinning dust grain possess a magnetic moment due to the Barnett effect\citep{Barnett1915}.
Its magnetization is
$M=\chi\Omega/\gamma$ \citep{Draine_ColdUniverse}, 
where $\chi$ is the magnetic susceptibility, $\Omega$ is the angular velocity,
$\gamma=g\mu_B/\hbar$ is the gyromagnetic ratio, and $\mu_B$ is the Bohr magneton. 
The $g$-factor is about 2 for electrons.

For regular paramagnetic material, the magnetic susceptibility follows Curie's Law \citep{Morrish2001}:
\begin{equation}
\label{eq:chi}
    \chi = \frac{n\mu^2}{3kT}=10^{-3}\hat{\chi}\left(\frac{T}{25\rm\, K}\right)^{-1},
\end{equation}
where $n$ is the number density of magnetic units, $\mu$ is the 
magnetic moment of each unit, and $\hat{\chi}$ is a dimensionless parameter that is on the order of
unity for regular paramagnetic materials \citep{Draine1996,Lazarian2007}.

The Larmor precession timescale is then:
\begin{equation}
    \begin{split}
    t_L = \frac{2\pi I\gamma}{\chi VB}
    =& 4.3 \times 10^{6} \mathrm{s}\times \hat{\chi}^{-1}
    \left(\frac{\rho_s}{3~\mathrm{g/cm^3}}\right)\\
    &\times \left( \frac{T_\mathrm{d}}{25~\mathrm{K}} \right) \left( \frac{B}{5~\mathrm{\mu G}} \right)^{-1} \left( \frac{a}{0.1~\mathrm{\mu m}} \right)^2.
    \end{split}
\end{equation}
We can see that this timescale is some six orders smaller than the gaseous damping timescale
for parameters appropriate for the diffuse ISM.  
But if we normalize the field strength to $5$ mG, which is typical for a PPD with $10^{-8}\rm\, M_\sun/yr$ accretion rate 
at tens of AU scale \citep{Bai2011}, and the grain size to $1$ mm, we get:
\begin{equation}
    \begin{split}
    t_L =& 4.3 \times 10^{11} \mathrm{s}\times \hat{\chi}^{-1}
    \left(\frac{\rho_s}{3~\mathrm{g/cm^3}}\right)\\
    &\times \left( \frac{T_\mathrm{d}}{25~\mathrm{K}} \right) \left( \frac{B}{5~\mathrm{mG}} \right)^{-1} \left( \frac{a}{1~\mathrm{mm}} \right)^2,
    \end{split}
    \label{eq:tL}
\end{equation}
which is about $10^3t_d$. 

\section{Superparamagnetic inclusions}
\label{sec:spi}

\subsection{Basic picture}
Superparamagnetic inclusions (SPIs) are small (nano-sized) particles of ferromagnetic 
material\footnote{In this work, we don't distinguish ferromagnetic material with ferrimagnetic 
material, or even speromagnetic material. They all behave as macro-spins, as discussed in the 
text, but maybe with different number of effective Bohr magneton per atom ($p$ in Eq.\eqref{eq:chisp}).}.
Within one such particle, all the atoms are spontaneously magnetized and behave like a single large 
magnetic moment, the so-called ``macro-spin" \citep{Bean1959}. They are not big enough to create 
domain walls yet, and are usually referred to as ``single-domain particles".
In the absence of external magnetic fields, these macro-spins are randomly oriented
and behave like paramagnetic materials as an ensemble.

Let's first consider the simplest case, where an SPI has no preferred direction for magnetization.
This isotropic case is mathematically identical to the regular paramagnetic case, and we have 
the bulk magnetic susceptibility of an ensemble of identical SPIs as \citep{JS1967}:
\begin{equation}
\chi_{sp}=\frac{\mathbb{N}\mu^2}{3kT},
\end{equation}
where $\mathbb{N}$ is the number density of SPIs inside the dust grain. Let $n_{tot}$ be the total
number density \ed{of atoms} and \ed{assume} a fraction of $f_{sp}$ atoms are magnetic atoms embedded in SPIs.
\ed{Let} $\mu$ \ed{be} the magnetic moment of the ``macro-spins".
If each SPI contains $N_{cl}$ magnetic atoms, we have $\mathbb{N}=f_{sp}n_{tot}/N_{cl}$, and
$\mu=N_{cl}p\mu_B$, where $p\mu_B$ is the averaged Bohr magneton for each magnetic atom. With these
we have (\citealt{Draine1996}):
\begin{equation}
\label{eq:chisp}
\begin{split}
\chi_{sp}=&\frac{f_{sp}n_{tot}N_{cl}(p\mu_B)^2}{3kT}\\
=&0.72\times10^{-2}N_{cl}f_{sp}\left(\frac{n_{tot}}{10^{23}\rm\,cm^{-3}}\right)
\left(\frac{p}{3}\right)^2\left(\frac{T}{25\rm\,K}\right)^{-1}.
\end{split}
\end{equation}
\ed{Compared with Eq.~\eqref{eq:chi}, we have:
\begin{equation}
\label{eq:chihat}
\hat{\chi} = 7.2 N_{cl} f_{sp} (p/3)^2.
\end{equation}
}

In reality, SPIs will have a preferred direction for magnetization. 
For example, a prolate particle prefers to be magnetized along its long axis \citep{Bean1959}. 
This is called shape anisotropy. 
Another example is metallic iron, which has a cubic crystalline structure. Metallic iron has less energy
when magnetized along one of the principal axes of its crystalline structure 
(so-called ``easy axis", \citealt{Dai2017}).
This is called magnetocrystalline anisotropy. 

Even though the energy would be different in the presence of anisotropy, the susceptibility remains the 
same in thermal equilibrium states. 
Let $K$ be the so-called ``anisotropy constant", such that $KV$ is the energy needed to change the
direction of the magnetic moment.
With a simple prescription, \cite{Bean1959} showed that the ensemble-averaged magnetic susceptibility
remains the same for both of the two limiting cases, when $KV\gg kT$ and when $KV\ll kT$. Thus 
Eq.~\eqref{eq:chisp} works even for anisotropic SPIs.

\subsection{Size limit determined by the relaxation process}
\label{subsec:sizelimit}
\ed{\cite{Billas1994} showed that single-domain particles}
as small as $N_{cl}=30$ can show superparamagnetism. 
In this subsection, we will discuss what determines the maximum size for SPI, which is more 
important than lower limit.

\cite{Neel1949} first proposed that single-domain particles experience random
Brownian-like motions that can change the orientation of its magnetic moment
(see also \citealt{Bean1959}).
This happens on a timescale (the so-called ``N\'{e}el's relaxation timescale") of:
\begin{equation}
t_N \equiv t_0 \exp\left(\frac{KV}{kT}\right),
\end{equation}
where $t_0$ is called ``attempt timescale" typically on the order of $10^{-9}\rm\, s$.
Setting the $t_N$ equal to the timescale of interest ($\tau$), we can define a critical 
blocking volume as: 
\begin{equation}
\label{eq:vcrit}
V_\mathrm{cr}=\frac{kT}{K}\ln\left(\frac{\tau}{t_0}\right).
\end{equation}
The typical dynamical timescale of a $100$ AU sized PPD is about $1000$ yr,
\ed{which yields $\ln(\tau/t_0)\approx 45$}.
We can see that the critical volume strongly depends on the temperature, and the cluster size measured at room 
temperature doesn't apply directly to an astronomical environment, which hasn't been considered before.
\ed{The critical number of magnetic atoms can then be calculated, for iron-based ferromagnetic material, through:
\begin{equation}
N_\mathrm{cr} = \frac{\rho V_\mathrm{cr}f_\mathrm{Fe}}{56m_p}=\frac{\rho f_\mathrm{Fe}kT}{56m_pK}\ln\left(\frac{\tau}{t_0}\right),
\label{eq:Ncr}
\end{equation}
where $f_\mathrm{Fe}$ is the mass fraction of iron atoms in the ferromagnetic
material constituting the SPIs, and $56m_p$ is the mass of an iron atom.}

\begin{figure}[!htp]
    \centering
    \includegraphics[width=0.5\textwidth]{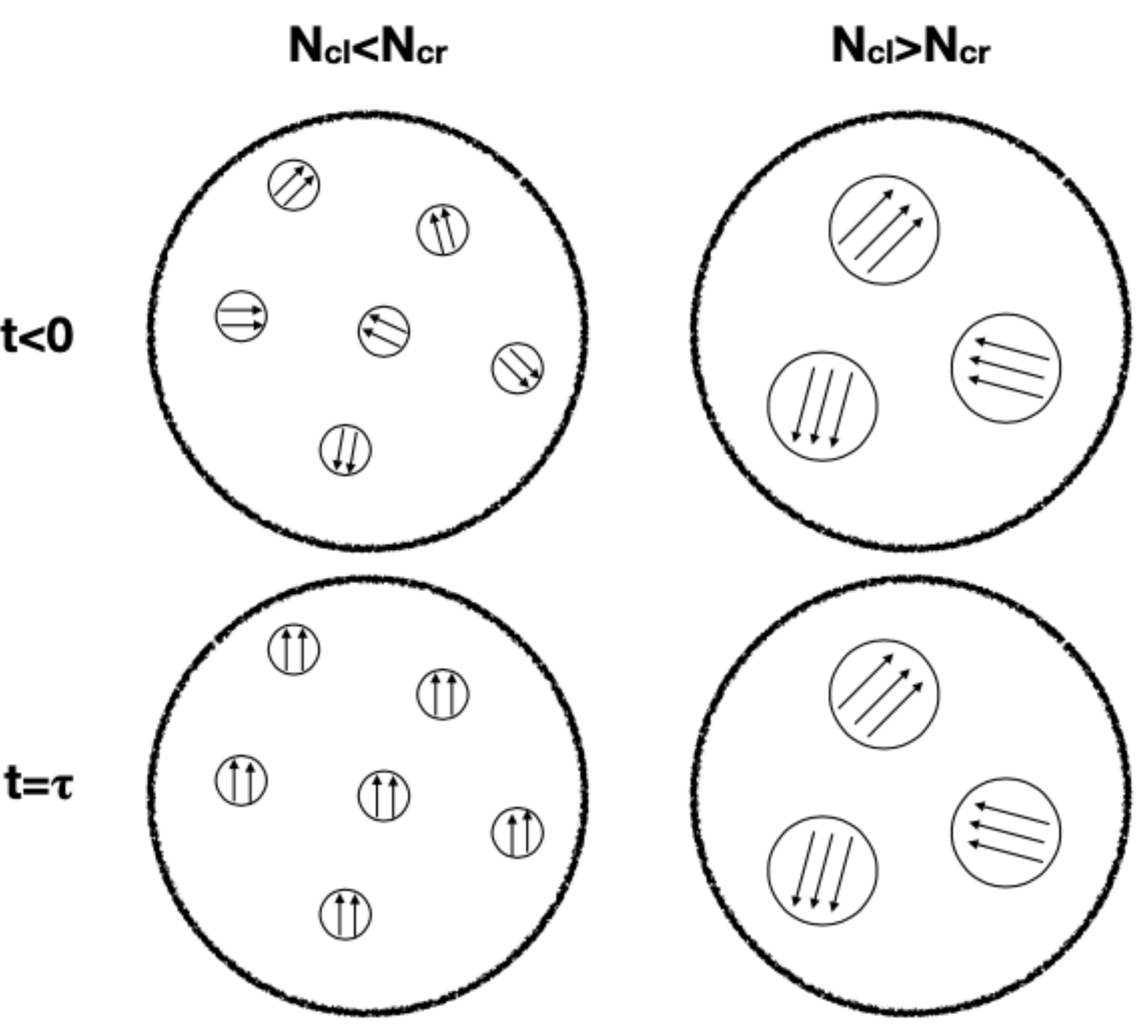}
    \caption{A schematic illustration of the magnetization of dust grains with SPIs of different sizes.
    The left column represents a dust grain with small SPIs. The right column represents a dust
    grain with big SPIs. The top row is their initial states before applying external magnetic fields. Starting
    from $t=0$, an external magnetic field going up is applied. The
    bottom row is the magnetization of these two dust grains at time $\tau$, \ed{the dynamical} timescale of interest.}
    \label{fig:scheme}
\end{figure}

The magnetization of a dust grain with SPIs is illustrated with a simplified model in 
Fig.~\ref{fig:scheme}. In this model, we consider two dust grains with uniformly sized SPIs. The left
one has smaller SPIs, whereas the right one contains bigger SPIs. At $t<0$, there is no external magnetic
field. The magnetic moment of each SPIs are randomly oriented such that both dust grains have no
bulk magnetization. At $t=0$, we turn on external magnetic fields, and observe the magnetization
of dust grains at $\tau$, the dynamical timescale of our interest. We will find that all the small SPIs
in the left dust grain turn into the external magnetic field direction (with thermal fluctuation). 
The resulting magnetic susceptibility of the dust grain is the superparamagnetic susceptibility $\chi_{sp}(N_{cl})$
in Eq.~\eqref{eq:chisp}. In contrast, the large SPIs in the right dust grain do not have enough time
to overcome the anisotropic energy barrier. Their magnetic moments are effectively ``blocked" and do 
not contribute to the magnetization of the dust grain. 
The dependence of the magnetic susceptibility on the size of SPI clusters can be approximated as 
(see Appendix~\ref{sec:analytical} for a more quantitative discussion): 

\begin{equation}
\chi = \left\{
\begin{array}{ll}
0, & N_{cl} > N_\mathrm{cr} \\
\chi_{sp}(N_{cl}), & N_{cl} < N_\mathrm{cr}
\end{array}\right..
\label{eq:chiblock}
\end{equation}

\ed{We can see that the maximum enhancement of the magnetic susceptibility for an ensemble of SPIs of
equal sizes is:}
\begin{equation}
\chi_\mathrm{max} = \chi_{sp}(N_\mathrm{cr}),
\label{eq:chimax}
\end{equation}
\ed{which is achieved when all SPIs in a dust grain are of the same size with 
$N_\mathrm{cr}$ magnetic atoms.}

\subsection{Distribution of SPIs}
So far we have only considered ensembles of SPIs with the same size. In order to understand 
the effects of a distribution of SPI sizes, we adopt a very simple power-law distribution with
index $-q$: $dn(N_{cl})/dN_{cl}=CN_{cl}^{-q}$, with $N_1<N_{cl}<N_2$ is the cluster size and 
$C$ is an arbitrary constant to be determined from the iron abundance. 

We are particularly interested in the scenario when $N_1<N_\mathrm{cr}<N_2$. With Eq.~\eqref{eq:chiblock},
the magnetic susceptibility of this ensemble of SPIs can be calculated as:
\begin{equation}
\begin{split}
\chi &= \frac{1}{B}\int_{N_1}^{N_2}\frac{dn(N_{cl})}{dN_{cl}} \mean{\mu_z} dN_{cl} \\
&= \int_{N_1}^{N_\mathrm{cr}} CN_{cl}^{-q} \frac{N_{cl}^2 p^2 \mu_B^2}{3kT} dN_{cl}\\
&= \frac{Cp^2\mu_B^2}{3(3-q)kT} \left.N_{cl}^{3-q} \right|_{N_1}^{N_\mathrm{cr}},
\end{split}
\end{equation}
where $\mean{\mu_z}$ is the averaged magnetic moment along magnetic field direction (see Appendix for more 
detail). With some arithmetic of finding the constant $C$, we can express our results as:
\begin{equation}
\chi = \chi_\mathrm{max}\times \left\{
\begin{aligned}
&\frac{2-q}{3-q}\left(\frac{N_\mathrm{cr}}{N_2}\right)^{2-q}, &q<2;\\
&\frac{q-2}{3-q}\left(\frac{N_1}{N_\mathrm{cr}}\right)^{q-2}, &2<q<3.\\
\end{aligned}
\right.
\end{equation}
We can see that the end magnetic susceptibility of this ensemble is always \ed{smaller than $\chi_\mathrm{max}$.}
The reduction factor is roughly the ratio of
the number density of iron atoms within SPIs with sizes close to $N_\mathrm{cr}$ to the total number
density of iron atoms within SPIs.
Because of this reduction factor, \ed{reaching the maximum value of $\chi_{sp}=\chi_\mathrm{max})$ implicitly assumes that all SPIs have sizes close to the critical size $N_\mathrm{cr}$.} 

\section{Estimate the critical size of SPIs}
\label{sec:spisize}

\ed{The critical size of an SPI is an important quantity for determining how big of an enhancement it will have 
on the magnetic susceptibility of the host dust grain.  In this section we perform estimates of SPI critical
sizes, in terms of the number of magnetic atoms $N_\mathrm{cr}$, for various materials that might plausibly 
be contained in astrophysical dust grains.}

As seen in Eq.~\eqref{eq:vcrit}, the anisotropy constant $K$ determines the critical
volume, when the temperature is fixed. In nature, there are two most important contributions to 
$K$: the shape anisotropy and the magnetocrystalline anisotropy. In this work, we will ignore
the first one and assume spherical SPIs. Including \ed{the shape anisotropy} will 
increase the anisotropy constant and decrease the size estimates given below. 
In other words, \ed{our} estimates \ed{are} conservative upper limits.

The critical volume is defined in Eq.~\eqref{eq:vcrit} and values at $25$ K are reported here. 
\ed{We assume room temperature densities of stoichiometric materials. }
Besides the \ed{critical} number of magnetic atoms in one SPI ($N_\mathrm{cr}$), \ed{we also report the 
magnetic susceptibility $\hat{\chi}_\mathrm{max}$, taking $f_{sp}=0.1$, and $n_{tot}=10^{23}\rm\, cm^{-3}$.}

\subsection{$\rm Fe_3O_4$ (magnetite)}
\ed{At low temperature ($T<120$ K), magnetite has a monoclinic structure \citep{Iizumi1982}. }
The magnetic anisotropy energy was determined by \cite{Abe1976} as:
\begin{equation}
E_a = K_a\alpha_a^2 + K_b\alpha_b^2+K_{aa}\alpha_a^4+K_{bb}\alpha_b^4+K_{ab}\alpha_a^2\alpha_b^2-K_u\alpha_{111}^2,
\end{equation}
with $K_a=25.2,\, K_b=3.7,\, K_u=2.1,\, K_{aa}=1.8,\, K_{bb}=2.4,\,$ and $K_{ab}=7.0$ in $10^5\rm\, erg/cm^3$,
and all $\alpha$'s are directional cosines (see \citealt{Abe1976} for their definitions).
The easy axes are along $(001)$ and $(00\bar{1})$ directions. The magnetization can change from one easy axis 
to another through saddle points along $(010)$ or $(0\bar{1}0)$ directions, and \ed{the energy barrier is 
$K\approx 6.1\times 10^5\rm\, erg/cm^3$. }
This translates to a critical volume as
$V_\mathrm{cr}=2.5\times 10^{-19}\rm\, cm^3$.
\cite{LI20071556} found a spin density as $3.54\mu_B$ per formula at $10$ K along [100] 
direction, which we will use in this paper. For every iron atom, we have $p=1.18$ Bohr magnetons.
Taking $\rho=5.17\rm\, g/cm^3$, we get $N_\mathrm{cr}=1.0\times10^4$ 
and $\hat{\chi}_\mathrm{max}=1.1\times 10^3$.

\subsection{$\gamma-\rm Fe_2O_3$ (maghemite)}
$\gamma-\rm Fe_2O_3$ (maghemite) is ferrimagnetic iron oxide with similar structures to magnetite at room
temperature. Its formula is often supposed to be $\rm (Fe^{3+})_A(Fe^{3+}_{5/3}\square_{1/3})_BO_4$, 
where $\square$ represents a vacancy. \ed{A perfect
crystal has $3.33\mu_B$ per formula. }
Hence we have $p=1.25$. 
In bulk samples, the moments were usually found to be about $87\sim94\%$ of this
value \citep{Coey1972}. 
\cite{Pisane2017} fitted the effective magnetic anisotropy as a function \ed{of particle 
size} with three terms. \ed{The leading term (and \ed{the} dominating term for big particles)
corresponds to $K=1.9\times 10^{5}\rm\, erg/cm^3$.}
This translates into a critical volume of $V_\mathrm{cr}=8.2\times 10^{-19}
\rm\, cm^3$.
Taking $\rho=4.9\rm\, g/cm^3$, we have $N_\mathrm{cr}=3.0\times10^4$,
and $\hat{\chi}_\mathrm{max}=3.7\times10^3$.

\subsection{Metallic iron}
The leading term in the magnetic anisotropy energy of metallic iron has the form of
$K_1(\alpha_1^2\alpha_2^2+\alpha_2^2\alpha_3^2+\alpha_1^2\alpha_3^2)$, where $\alpha_i$, $i=1,2,3$ are directional 
cosines. At temperatures below $100$ K, $K_1\approx 5.4\times 10^{5}\rm\, erg/cm^3$ \citep{Dai2017}.
\ed{The energy barrier to change the magnetization} from one easy axis to another \ed{is}
$K=(1/4)K_1=1.35\times10^{5}\rm\, erg/cm^3$.
Hence $V_\mathrm{cr}=1.15\times10^{-18}\rm\, cm^3$.
We will follow \cite{Draine1996} and take $p=3$, which was inspired by \cite{Billas1994}'s
work showing that small clusters of iron have $3\mu_B$ per atom. 
Taking $\rho=7.87\rm\,g/cm^3$, we have $N_\mathrm{cr}=9.7\times 10^4$ and $\hat{\chi}_{\rm max}=7.0
\times10^4$.

\subsection{Other forms of iron and summary}
Hematite ($\alpha-\rm Fe_2O_3$) is another possible form of iron, which is more 
stable than maghemite ($\gamma-\rm Fe_2O_3$) discussed above. 
It is, however, antiferromagnetic, and perfect crystal can be essentially considered 
as non-magnetic. In reality, some defects may exist to contribute to the magnetization but it should
be negligible comparing with other ferromagnetic materials discussed above. The same goes for FeO. 
For a summary of magnetic properties of the iron oxides, we refer interested readers to
\cite{IronOxides}, especially their Table~6.2.
\ed{We do not consider sulfuric iron in this work.}

All results discussed above are summarized in Table~\ref{tbl:ncl}. We can see the maximum
cluster size and the enhancement of magnetic susceptibility strongly depends on the form of iron.
Even though $\hat{\chi}_\mathrm{max}$ on the order of $10^4$ is still possible with metallic iron,
it is less likely to exist in real dust grains as it can easily be oxidated. We suggest that $10^3$ is 
a more realistic enhancement factor $\hat{\chi}_\mathrm{max}$ on the magnetic susceptibility through SPIs.

\begin{table*}[!htb]
\centering
\begin{tabular}{c|ccccc}
\hline
Material & $K$ ($\rm erg/cm^3$) & $p$ & $V_\mathrm{cr}$ ($\rm cm^{3}$) & $N_\mathrm{cr}$ & $\hat{\chi}_{\rm max}$\\
\hline
$\rm Fe_3O_4$ & $6.1\times 10^5$ & $1.18$ & $2.5\times 10^{-19}$ & $1.0\times 10^3$ & $1.1\times10^3$\\
$\rm \gamma-Fe_2O_3$ & $1.9\times 10^5$ & $1.25$ & $8.2\times 10^{-19}$ & $3.0\times 10^4$ & $3.7\times 10^3$\\
Fe & $1.35\times 10^5$ & $3$ & $1.15\times 10^{-18}$ & $9.7\times 10^4$ & $7.0\times 10^4$\\
\hline
\end{tabular}
\caption{Anisotropy constant $K$, effective number of Bohr magnetons per iron atom $p$, critical 
volume $V_\mathrm{cr}$ (Eq.~\eqref{eq:vcrit}), \ed{critical cluster size $N_\mathrm{cr}$, and
reduced magnetic susceptibility $\hat{\chi}_{\rm max}$}
for various iron bearing ferromagnetic materials that may exist in dust grains. See text for
discussion and references. }
\label{tbl:ncl}
\end{table*}

\section{Magnetic alignment in PPDs}
\label{sec:disk}
In Sec.~\ref{sec:basics}, we gave rough estimates of timescales relevant for magnetic alignment and compared 
them to motivate this study. In this section, we perform a more detailed study on whether magnetic alignment is 
feasible in a fiducial PPD model.

\subsection{Disk model}
As an illustration, we adopt the well-known \cite{CG1997} model. 
It is a passive disk with Minimum Mass Solar Nebula density profile
$\Sigma=(r/\mathrm{AU})^{-3/2}\Sigma_0$, with $\Sigma_0=10^3\rm\, g/cm^2$ \citep{Weidenschilling1977}.
It has a superheated surface layer and a cooler interior region, where most grown mm-sized dust grains
reside.
The temperature in the interior region and the scale height in the model under both hydrostatic equilibrium 
and radiative equilibrium were fitted as
\begin{equation}
T = 
\begin{dcases}
150\,\mathrm{K}\times \left(\frac{r}{1\rm\, AU}\right)^{-3/7} & 0.4\,\mathrm{AU}<r<84\,\mathrm{AU}\\
21\,\mathrm{K}& 84<r<100\,\mathrm{AU}\\
\end{dcases},
\end{equation}
and
\begin{equation}
\frac{H}{r} = 
\begin{dcases}
0.17\left(\frac{r}{1\rm\, AU}\right)^{2/7} & 0.4\,\mathrm{AU}<r<84\,\mathrm{AU}\\
0.59\left(\frac{r}{84\rm\, AU}\right)^{1/2}& 84<r<100\,\mathrm{AU}\\
\end{dcases}.
\label{eq:Tprofile}
\end{equation}
At a given radius, the midplane density is used to calculate timescales, which is $\Sigma/\sqrt{2 \pi H}$.
We adopt a mean molecular weight of $2.3$ and assume the temperature to be the same for gas and dust. 

For the magnetic field structure, we adopt the estimate from \cite{Bai2011}:
\begin{equation}
B=1.0\,\mathrm{G}\times \left(\frac{\dot{M}}{10^{-8}\rm\,M_\sun/yr}\right)^{1/2}\left(\frac{r}{1\rm\,AU}\right)^{-11/8},
\label{eq:Bmod}
\end{equation}
where $\dot{M}$ is the mass accretion rate, which is assumed to be $10^{-8}\rm\, M_\sun/yr$,
typical for classical T Tauri stars (see \citealt{Dutrey2014} and references therein).

\subsection{Timescale comparison}
As discussed in Sec.~\ref{subsec:criteria}, we are specially interested in 
two conditions: $t_L=t_d$, 
and $t_L=0.1t_d$. These two conditions define three different regimes:
\begin{itemize}
    \item $t_L>t_d$: No magnetic alignment.
    \item $0.1t_d<t_L<t_d$: More complicated dynamical study needed. 
    \item $t_L<0.1t_d$: Magnetic alignment is possible with fast Larmor precession.
\end{itemize}

\begin{figure}[!htp]
    \centering
    \includegraphics[width=0.5\textwidth]{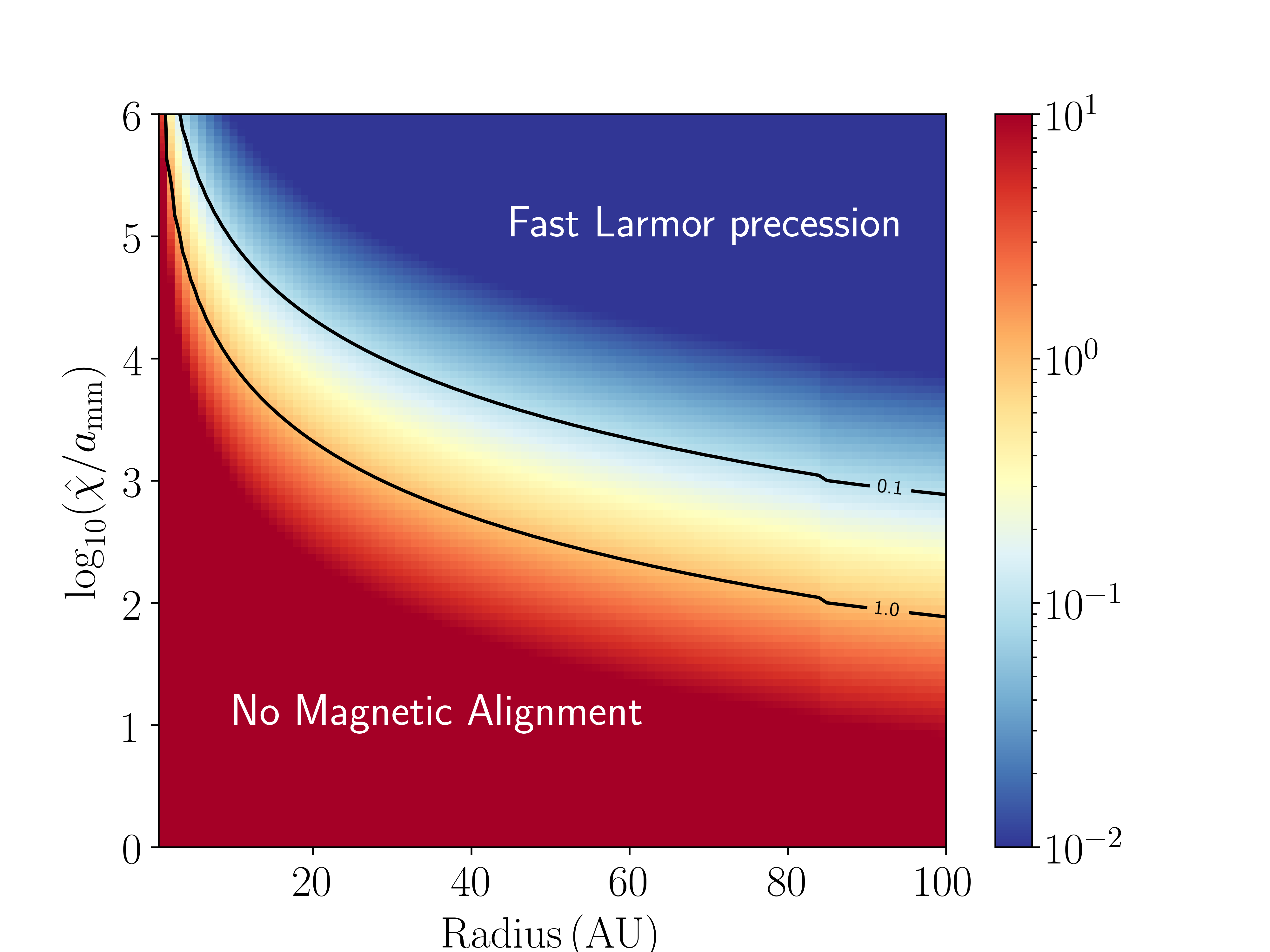}
    \caption{The ratio $t_L/t_d$ as a function of the radius and $\hat{\chi}/a_\mathrm{mm}$, with 
    $a_\mathrm{mm}\equiv a/(1\rm\, mm)$. The $t_L/t_d=0.1$ and $t_L/t_d=1$ contours are plotted
    to aid the interpretation.}
    \label{fig:tratio}
\end{figure}

From Eq.\eqref{eq:tL} and \eqref{eq:td}:
\begin{equation}
\begin{split}
\frac{t_L}{t_d}=& 1.65\times10^3 \left(\frac{\hat{\chi}}{a_\mathrm{mm}}\right)^{-1}
\left(\frac{T}{25\rm\, K}\right)^{3/2}\\
&\times \left(\frac{n_g}{5\times10^9\rm\, cm^{-3}}\right)\left(\frac{B}{5\rm\, mG}\right)^{-1},
\end{split}
\label{eq:tratio}
\end{equation}
where $a_\mathrm{mm}\equiv (a/1\,\mathrm{mm})$. We can see that $\hat{\chi}$ and grain size $a$ are
degenerate. 
Due to this degeneracy, the timescale ratio is calculated as a function of radius in the PPD, and
the dimensionless \ed{factor} $\hat{\chi}/a_\mathrm{mm}$. The results are plotted in Fig.~\ref{fig:tratio}.

First of all, we can see that grains in the inner disk are harder to align compared with those in the outer
disk. This is because the gas density power low index ($-2.8$ in our model within $84$
AU) is usually more negative than the magnetic fields power low index ($-1.4$ in our model).
As we decrease the radius, the gaseous damping timescale increases faster than the 
Larmor precession timescale. 
\ed{In order to achieve fast Larmor precession ($t_L<0.1t_d$)} at a radius of 30 AU, which is the typical resolution of ALMA polarization
observations of the nearest star forming regions, \ed{we need roughly} $\hat{\chi}/a_\mathrm{mm} > \sim 10^4$.

\begin{figure}[!htp]
    \centering
    \includegraphics[width=0.5\textwidth]{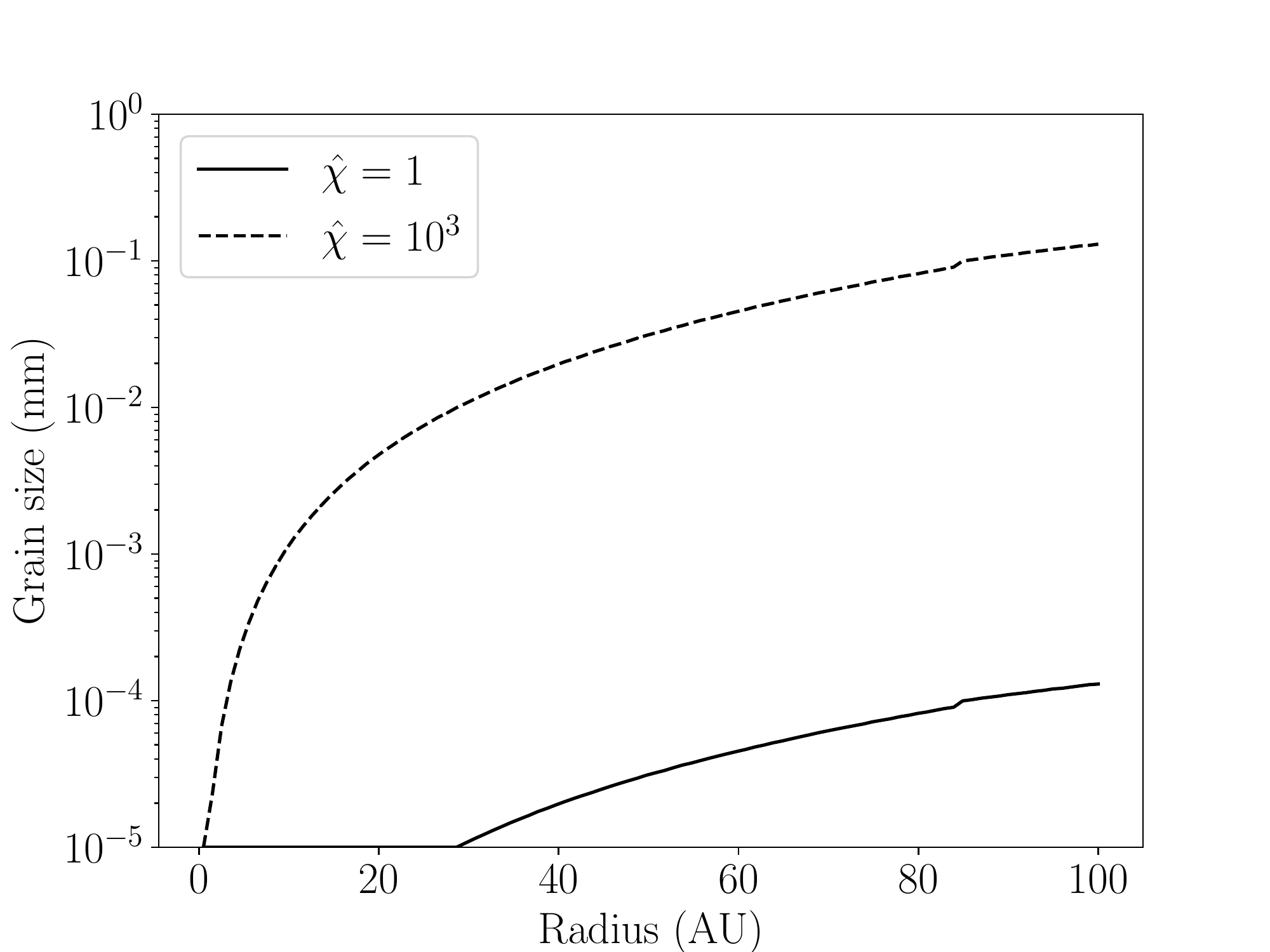}
    \caption{The largest grain sizes with fast Larmor precession ($t_L/t_d<0.1$) at each radius in the
    disk. Both regular paramagnetic dust grains ($\hat{\chi}=1$) and grains with substantial SPIs 
    ($\hat{\chi}=10^3$) are plotted. }
    \label{fig:grainsize}
\end{figure}

This result can be easily translated \ed{into} more meaningful statements after fixing $\hat{\chi}$. 
Fig.~\ref{fig:grainsize} shows the largest grain sizes with fast Larmor precession ($t_L/t_d<0.1$) 
at each radius in the disk. 
If one take 
$\hat{\chi}=10^3$, the upper limit we suggested in Sec.~\ref{sec:spisize}, we get
$a=100\rm\, \mu m$. This means that grains with sizes of $100\rm\, \mu m$ or bigger 
are unlikely to be aligned with magnetic field within $30$ AU of the central star, 
even with the aid of SPI. \ed{The situation is even worse for regular paramagnetic 
material ($\hat{\chi}=1$), where we need $a>0.1\rm\, \mu m$ to have fast Larmor precession.}

The situation is a lot better in the outer disk at scales larger than $100$ AU. We
will only need $\hat{\chi}/a_\mathrm{mm}\sim 10^3$. Alignment of $1$ mm dust grains
becomes marginally possible if they contain large SPI clusters ($\hat{\chi}>\sim10^3$). 

\section{Discussion}
\label{sec:discussion}

\subsection{Caveats and uncertainties in this work}
\ed{The above discussions have relied on a number of simplifying assumptions.  In this section we enumerate 
these assumptions and discuss the implications that relaxing them could have for our results.}

\subsubsection{Suprathermal rotation}
\ed{Our first assumption is that the dust grain rotational motions are distributed thermally, meaning that we
have effectively ignored the possibility of suprathermal rotation.}
\cite{Purcell1979} first pointed out that the dust grains can be spun up
to suprathermal motion (rotation energy much larger than thermal energy) \ed{by} torques \ed{arising} from the
formation of molecular Hydrogen \ed{on} the surface of the dust grains. 
A suprathermally rotating grain \ed{would} have a \ed{larger} gaseous damping timescale, which \ed{would make}
grain alignment \ed{easier}.
However, this Hydrogen formation torque is unlikely to work for large millimeter-sized dust grains, 
since the rotation energy decreases with increasing number of Hydrogen formation sites. 
In \ed{PPDs}, \ed{more plausible mechanisms for producing suprathermal rotation in millimeter-sized dust grains 
are radiative alignment torques} (RAT; \citealt{LH2007}) and mechanical torques from differential motion with 
\ed{the surrounding} gas (MAT; \citealt{Hoang2018}).
Indeed, both \cite{LH2007} and \cite{Hoang2018} showed that grains can be aligned towards the so-called 
``high-J" attractors with angular momenta \ed{that are factors of several tens to hundreds of times larger 
than} the thermal angular momentum.
However, \cite{Hoang2018} also showed that the mechanical alignment torque goes as $1/\sqrt{N_\mathrm{facet}}$, with 
$N_\mathrm{facet}$ being the number of facets \ed{on} the grain surface. This cancellation effect may 
\ed{prevent} the MAT \ed{from working} for large millimeter-sized grains. 
\ed{Such grains may also have limited helicity, which is crucial for both the RAT and the MAT to operate, 
so it is possible that this cancellation effect applies to RAT as well.}  
\ed{More} detailed study on suprathermal rotation of large millimeter-sized grains in dense environment is 
needed to \ed{better understand how they might modify our results.}

\subsubsection{Alignment Condition}
\ed{Our second assumption is that of fast Larmor precession, coupled with the lack of a precise understanding 
regarding what constitutes ``fast enough'' precession.}  
Previous work on magnetic grain alignment has focused on two limiting cases.
In \ed{the first limit, that of fast Larmor precession ($t_L \ll t_d$), torques are taken 
to have values that are averaged over the precession cycle prior to investigating grain 
dynamics (e.g., B-RAT work in \citealt{LH2007} and B-MAT work in \citealt{Hoang2018}).}
In the other limit, \ed{that of slow Larmor precession ($t_L \gg t_d$),} the effects of magnetic 
fields are completely ignored \ed{(e.g. k-RAT work in \citealt{LH2007} and k-MAT work in \citealt{Hoang2018})}. 
\ed{However, as we can see }from Fig.~\ref{fig:tratio}, it is not unreasonable \ed{to 
expect that} the ratio $t_L/t_d$ \ed{may} fall in the range $0.1 - 1$ \ed{in PPD 
environments, in which case} we can neither assume precession-averaged torques nor ignore 
the magnetic field completely.  
Whether dust grains can be aligned magnetically in this regime 
is thus not \ed{currently} clear, and this uncertainty limits the \ed{predictive} power of 
magnetic alignment theories. Solving this problem \ed{will require} studying the grain dynamics in three \ed{dimensions (i.e., without averaging over Larmor precession) for each potential alignment mechanism. }

\subsubsection{Form of iron}
\ed{Our third assumption is the iron abundance.} In deriving
$\hat{\chi}_\mathrm{max}$, we took $f_{sp}=0.1$.
\ed{\cite{Draine1996} suggested that about $10\%$ of atoms in a grain are iron atoms.
$f_{sp}=0.1$ means all iron atoms are in the form of SPIs, and no iron atoms are in other
forms, such as silicate.}
This optimistic assumption can easily be wrong and reduce $\hat{\chi}_\mathrm{max}$ 
by \ed{one} order of magnitude \ed{or more}, making magnetic alignment harder. 
We also see that the maximum cluster size depends on the detailed form of
iron. Even for ferromagnetic materials, the \ed{critical} cluster size can vary
by orders of magnitude. A chemical study of dust compositions may help
resolve this uncertainty. 

\subsubsection{Disk model}
\ed{Our fourth assumption is the specific form of the PPD disk model.  The MMSN model, by 
construction, contains only the minimum amount of mass required to form our solar system. 
It is thus likely to have less mass than a real PPD. Increasing the mass and density of a
PPD would increase the gaseous damping timescale, which would make magnetic grain alignment 
more difficult.}

\subsection{Effects of temperature dependence on cluster sizes}
In Sec.~\ref{sec:disk}, we used $\hat{\chi}$ as a free parameter to discuss
the possibility of magnetic alignment in PPDs, even though the $\hat{\chi}_\mathrm{max}$
has a temperature dependence. In this \ed{section}, we discuss the 
\ed{implications} of this temperature dependence. 

One may be tempted to set $\hat{\chi}$ at each radius in the disk to the local 
$\hat{\chi}_\mathrm{max}$ defined by its local temperature. However, doing so would 
mean that \ed{the SPI sizes in dust grains are changing} at each
location \ed{in the disk, in such a way as to maximize} the effects of SPM. 
This \ed{behavior} is not physical, \ed{both because} dust 
grains at different radii \ed{should} have similar origins and 
\ed{because the} size of SPIs in \ed{any} given dust grain \ed{are} unlikely to change
as it migrates to other locations of the disk. 

However, the temperature dependence of maximum cluster sizes may still have an
impact on the magnetic susceptibility \ed{of dust grains as a function of disk radius}. 
If the dust grains contain SPIs \ed{with sizes larger than the critical $N_{cr}$ (see
Eq.~\ref{eq:Ncr}), then}
it is possible that these SPIs \ed{will be} blocked in the \ed{low-temperature regions} 
of the disk, while contributing \ed{heavily} to the magnetic susceptibility
in the \ed{high-temperature regions of the disk}. \ed{Such a} model would predict
a higher degree of magnetic alignment near the center of the disk. 
\ed{It} is \ed{interesting} \ed{to note that this behavior is} opposite \ed{that
expected from} the usual Curie's Law (Eq.~\ref{eq:chi}). 

\ed{In addition to its} explicit linear dependence on temperature, the critical volume
has another implicit temperature dependence through the anisotropic 
constant $K$. For metallic iron, this dependence can be \ed{safely} ignored as it 
changes very slowly and smoothly \ed{within} the temperature range $0$--$300$ K 
(\ed{decreasing} by roughly only $15\%$ \ed{across the range}; \citealt{Dai2017}). 
For magnetite ($\rm Fe_3O_4$), however, \ed{the situation} is \ed{considerably more} complicated. 
\ed{The first complication is that} magnetite undergoes \ed{a}
Verwey transition \citep{Verwey1939,Walz2002} and \ed{thus} changes crystal structure 
\ed{at a temperature of} around $120$ K \citep{Iizumi1982};
\ed{the structure changes from cubic spinel above this temperature to monoclinic below it}.
Our results for magnetite above \ed{(Sec.~\ref{sec:spisize}) would thus} need to be
revisited \ed{for} environments with temperature higher than $120$ K.
\ed{A second complication is that magnetite} has so-called ``isotropic point" near $130$ K. Around this temperature, the cubic 
anisotropy constant changes signs, \ed{such that} easy axis changes from the cubic diagonals ($T>130$ K) to
cubic edges ($T<130$ K); in the \ed{immediate} vicinity of $T\sim 130$ K, the anisotropy constant is close
to zero. \ed{A near-zero anisotropy constant permits} an arbitrarily large \ed{SPI cluster size}\footnote{\ed{The cluster size} may still be limited by the
domain size -- which is usually larger than the blocking size and \ed{which} is ignored in this work -- as well as other 
sources of anisotropy such as the shape anisotropy.}. 
If dust grains contain \ed{large} clusters of magnetite, they may be blocked at 
temperatures other than $\sim130$ K, \ed{in which case our} model would predict a ring of magnetically aligned 
dust grains \ed{only} in the \ed{region of the} disk \ed{where} $T\approx 130$ K.
However, \ed{we note that} for typical PPDs such a high temperature is reached only on the AU scale (see 
Eq.~\ref{eq:Tprofile}), which is unlikely to be resolved by ALMA in polarized emission.

\subsection{Applications to observed disks}

\subsubsection{Small grains near the surface of AB Aur disk}
\cite{Li2016} used mid-infrared polarimetry to observe the disk around AB Aur \ed{at a 
wavelength of} about $10.3\rm\, \mu m$. The polarization was \ed{interpreted as arising from}
dust grains aligned with poloidal magnetic fields.

At this wavelength, dust grains with sizes on the order of $\sim \rm \mu m$ are most
important for emission. \ed{Such} small dust grains \ed{are} lifted \ed{more easily} by 
turbulence \ed{and thus reach} higher \ed{above the disk midplane than} (sub)millimeter grains. 
Both the dust grains and the \ed{their} environment are \ed{thus} different from \ed{what we
have discussed in Sec.~\ref{sec:disk}}.
\ed{For the sake of example, suppose} the grains are lifted to \ed{twice the} scale height, 
\ed{where} the gas density \ed{is an} order of magnitude smaller \ed{than in the disk midplane}. The magnetic field is also \ed{likely to be} different in this region,
but we will \ed{assume} the same value as in Eq.~\eqref{eq:Bmod}. \ed{Given these conditions}, we derive the
criterion for fast Larmor precession to be $\hat{\chi}/a_\mathrm{mm}=10^3$ at $30$ AU 
(\citealt{Li2016} has a 50 AU \ed{resolution}). For $\mu$m-sized grains, SPIs \ed{are thus
unnecessary to achieve} magnetic alignment \ed{because} $\hat{\chi}=1$ is \ed{sufficient} to 
guarantee fast Larmor precession. Mid-IR polarimetry probing the 
small dust grains at the disk surface \ed{should therefore be capable of studying} magnetic
field structures in PPD. 

\subsubsection{Class 0 systems}
Class 0 disks \ed{represent} the earliest \ed{stages of planet} formation. 
They are likely to have stronger magnetic fields \citep{Yen2017} and may be more massive \ed{than their later-stage counterparts} \citep{Tobin2020}.
If we adopt a representative accretion rate, $10^{-6}\rm\, M_\sun /yr$, we would increase the
strength of magnetic field by one order of magnitude, compared with the value adopted in 
Eq.~\eqref{eq:Bmod}. It is possible that Class 0 disks are also more massive than Class I/II disks. 
For example, \cite{Tobin2020} reported the mass of Class 0 and Class I disks has mean dust mass 
of $25.9M_\oplus$ and $14.9M_\oplus$, respectively. 
If we assume the mass of the Class 0 disk is the same \ed{as the one adopted in Sec.~\ref{sec:disk}} -- hence the density is also the same -- 
\ed{then applying the} fast Larmor precession \ed{criterion} yields $\hat{\chi}/a_\mathrm{mm}=10^3$. 
\cite{Valdivia2019} infer that the dust grains \ed{in Class 0 sources} have grown to at least $10\rm\, \mu $m.
\ed{Such dust grains} require at least $\hat{\chi}=10$ \ed{to have fast Larmor precession},
\ed{which requires minimal} SPI enhancement to achieve. 
At the same time, $10\rm\, \mu$m dust grains are less efficient at 
producing polarization through  
self-scattering at (sub)millimeter wavelengths, \ed{avoiding a known confounding effect 
for studying magnetic fields} \citep{Kataoka2015}. 
Class 0 disks \ed{are thus reasonable} targets for detecting magnetic fields through
\ed{spatially resolved} (sub)millimeter polarimetry.

\section{Summary}
\label{sec:summary}
In this paper, we discussed the feasibility of magnetic alignment of dust grains with superparamagnetic
inclusions in protoplanetary disks. The major results are summarized as follow:
\begin{enumerate}
\item Under the N\'{e}el's relaxation theory, we show that there exists a critical size 
of SPIs \ed{within dust grains}. \ed{SPIs larger} than this \ed{critical}
size cannot respond to external magnetic fields and do not contribute to the magnetic susceptibility \ed{of the grain}.
\ed{There is thus a} maximum enhancement \ed{that SPIs can provide to a grain's}
magnetic susceptibility, and \ed{therefore a}
corresponding maximum reduction factor for the Larmor precession timescale \ed{of the grain}.
\item 
\ed{We explore the effect on magnetic susceptibility for a dust grain containing an ensemble of SPIs having a power-law size distribution. }
We find that if the ensemble contains SPIs bigger than the threshold $N_\mathrm{cr}$, \ed{the 
grain as a whole will be unlikely to achieve the maximum magnetic susceptibility $\chi_\mathrm{max}$ (Eq.~\ref{eq:chimax})
without fine-tuning (i.e., all SPIs in the dust grain would need to have sizes close to the 
maximum cluster size). }
This is because SPIs larger than \ed{the critical} size \ed{do not respond to external magnetic fields} and do not contribute to the 
ensemble magnetic susceptibility \ed{(Eq.~\ref{eq:chiblock})}.

\item \ed{We estimate the maximum sizes for SPIs composed of several plausible ferromagnetic
materials,} given their magnetic anisotropy constant at the temperature of astronomical 
interest ($25$ K). Our results are tabulated in Table~\ref{tbl:ncl}. We suggest that $10^3$ is a more 
realistic upper limit for the SPI enhancement \ed{factor $\hat{\chi}$}. 
This \ed{value} is two orders of magnitude smaller than \ed{that obtained from} previous work,
\ed{implying that} magnetic alignment \ed{is} more difficult than previously thought. 

\item 
We explore the feasibility of magnetic grain alignment in the disk midplane of a MMSN model for a PPD. 
We find that (1) magnetic alignment is impossible \ed{unless} SPIs \ed{are ubiquitous} throughout the disk, even 
\ed{if the dust} grains \ed{are} as 
small as $10\rm\, \mu m$; (2) it is difficult to align grains \ed{larger than} 
100 $\rm \mu m$ even with SPIs, \ed{particularly at small orbital radii where the high
gas density leads to short damping timescales.}
\end{enumerate}
We conclude that large millimeter-sized dust grains in the midplane of PPDs are unlikely to be aligned with the
\ed{ambient} magnetic fields (c.f.~Fig.~\ref{fig:grainsize}). 
\ed{An important implication of this finding is that observations that are primarily sensitive 
to the emission from this population of grains -- such as (sub)millimeter-wavelength
polarimetric observations of Class I/II PPDs -- are unlikely to be tracing the disk 
magnetic field structure.}
We suggest \ed{instead} that \ed{observations of} disk surfaces -- probed by mid-infrared
polarimetry -- and the early Class 0 PPDs are \ed{better-suited} to studying the magnetic fields
structures \ed{in PPDs}.

\section*{Acknowledgements}

We thank the anonymous referee for detailed and constructive comments that greatly improved the manuscript.
We thank Xuening Bai, Zhi-Yun Li, Daniel Harsono, and Vincent Guillet for fruitful discussions.
We thank Zheng Liu, Zhiyuan Yao and Shuai Yin for discussions over the physics behind 
the superparamagnetism. We thank Dominic Pesce, Zhi-Yun Li, and Xuening Bai for comments and 
suggestions that helped to improve the manuscript. 

\appendix
\section{A simple model for SPI dynamics}
\label{sec:analytical}
In order to see how the magnetic susceptibility of an ensemble of identical SPIs changes as a function of 
their volume, we use a simplified model with three-axis assumption: all the SPIs have their easy axis
along one of an arbitrary set of Cartesian axes and they are split evenly among three axes. Also, we are interested
in the regime where $V>kT/K$, so the equilibrium state have averaged magnetic moment as $\mu^2B/kT$ along easy
axis and $0$ along perpendicular directions. Initially, all the SPIs have their magnetic moment randomly
distributed, so that there are equal numbers of $\mu$'s along the six directions ($\pm x, \pm y,$ and $\pm z$).

Now let's apply magnetic field $B$ along $z$ axis at $t=0$.
Since the Barnett equivalent field (for thermal angular velocity of a $10\rm\, \mu m$ dust grain)
$H_B=\Omega/\gamma\sim~10^{-7}\rm\,Oersted$ and the corresponding magnetic energy is at most on the order of 
$10^5\mu_BH_B/k\sim 10^{-7}\rm\, K$, which is very small compared with the thermal energy (on the order of
$10$ K). We conclude that the magnetic field cannot overcome the anisotropy energy on its own, 
which is even bigger than the thermal energy. We need the Brownian-like process N\'{e}el proposed 
that each magnetic 
moment tries to change its orientation every $t_0$. Since the SPIs with easy axis along $x$ or $y$ 
direction cannot response to the magnetic field along $z$ direction, we focus on those with easy axis along
$z$ axis, and define $f_+$ and $f_-$ as fraction of particles along $z+$ and $z-$, respectively. 
The energies for these two states are $-\mu B$ and $\mu B$, but there is an energy barrier of $KV$ if the
magnetic moment tries to switch to the opposite state. As such, the probability of one successful transition
is:
\begin{equation}
\left\{
\begin{split}
P(z+\to z-)=\exp\left(-\frac{KV+\mu B}{kT}\right)\\
P(z-\to z+)=\exp\left(-\frac{KV-\mu B}{kT}\right)
\end{split}
\right..
\end{equation}
The dynamical equation for $(f_+,f_-)$ is then:
\begin{equation}
\frac{d}{dt}\left(\begin{array}{c}f_+ \\ f_-\end{array}\right) = 
\frac{1}{t_0}\left(
\begin{array}{cc}
-\exp\left(-\frac{KV+\mu B}{kT}\right)& \exp\left(-\frac{KV-\mu B}{kT}\right)\\
\exp\left(-\frac{KV+\mu B}{kT}\right)& -\exp\left(-\frac{KV-\mu B}{kT}\right)\\
\end{array}\right)
\left(\begin{array}{c}f_+ \\ f_-\end{array}\right).
\end{equation}
The solution to this equation is:
\begin{equation}
\left(\begin{array}{c}f_+ \\ f_-\end{array}\right) = \exp\left(\frac{t}{t_N}\mathcal{M}\right)
\left(\begin{array}{c}f_{+,0} \\ f_{-,0}\end{array}\right),
\label{eq:solution}
\end{equation}
where $t_N$ is the N\'{e}el's relaxation timescale, 
and $\mathcal{M}$ is a matrix defined as 
$\mathcal{M}\equiv [[-e^{-\beta},e^{\beta}],[e^{-\beta},-e^{\beta}]]$, with $\beta\equiv\mu B/kT$.

It is possible to calculate the above matrix exponential explicitly with the aid
of the following matrix transformation:
\begin{equation}
\mathcal{M} = 
\mathcal{P}
\left(
\begin{array}{cc}
0 & 0 \\
0 & -2\cosh{\beta}\\
\end{array}\right)
\mathcal{P}^{-1},
\end{equation}
with $\mathcal{P}\equiv [[e^\beta, 1], [e^{-\beta}, -1]]$. Eq.~\eqref{eq:solution} can then be rewritten
as:
\begin{equation}
\left(\begin{array}{c}f_+ \\ f_-\end{array}\right) = 
\mathcal{P}\left(
\begin{array}{cc}
1 & 0 \\
0 & \exp\left(-\frac{2t}{t_N}\cosh{\beta}\right)\\
\end{array}\right)\mathcal{P}^{-1}
\left(\begin{array}{c}f_{+,0} \\ f_{-,0}\end{array}\right).
\label{eq:solution2}
\end{equation}
\ed{Now let's consider the} two limiting cases. For $t\gg t_N$, the exponential in 
Eq.~\eqref{eq:solution2} is basically $0$. We have:
\begin{equation}
\left(\begin{array}{c}f_+ \\ f_-\end{array}\right) = 
\frac{f_{+,0}+f_{-,0}}{2\cosh\beta}\left(\begin{array}{c}e^\beta \\ e^{-\beta}\end{array}\right).
\end{equation}
This means the end state is independent of initial state and the system has reached equilibrium 
(we always have $f_{+,0}+f_{-,0}=1$). 
In this case, the averaged magnetic moment along magnetic field direction is 
$\mean{\mu_z}= f_+\mu-f_-\mu =\mu\tanh\beta \approx \mu^2B/kT$. 
The magnetic susceptibility of this ensemble is thus the same as the superparamagnetic magnetic susceptibility
$\chi_{sp}$ in Eq.~\eqref{eq:chisp} (remember that there are two third of SPIs with perpendicular easy axes and do not contribute 
to the magnetic susceptibility).

For $t\ll t_N$, we have (this is easier to derive directly from Eq.~\ref{eq:solution}):
\begin{equation}
\left(\begin{array}{c}f_+ \\ f_-\end{array}\right) \approx
\left(\mathcal{I}+\frac{t}{t_N}\mathcal{M}\right)
\left(\begin{array}{c}f_{+,0} \\ f_{-,0}\end{array}\right)
=\left(\begin{array}{c}f_{+,0} \\ f_{-,0}\end{array}\right)
+\frac{t}{t_N}
\left(\begin{array}{c}-f_{+,0}e^{-\beta}+f_{-,0}e^\beta \\ f_{+,0}e^{-\beta}-f_{-,0}e^{\beta}\end{array}\right).
\end{equation}
From the end result, we can see that a fraction of $(t/t_N) e^{-\beta}$ SPIs have changed from
$+$ state to $-$ state, whereas a fraction of $(t/t_N) e^{\beta}$ SPIs have changed from $-$ state
to $+$ state. In this case, for an initial condition $f_{+,0}=f_{-,0}=0.5$, we get
$\mean{\mu_z} = f_+\mu-f_-\mu = (2t\mu/t_N)\sinh\beta\approx (2t/t_N)\mu^2B/kT$.

\begin{figure}[!htp]
    \centering
    \includegraphics[width=0.5\textwidth]{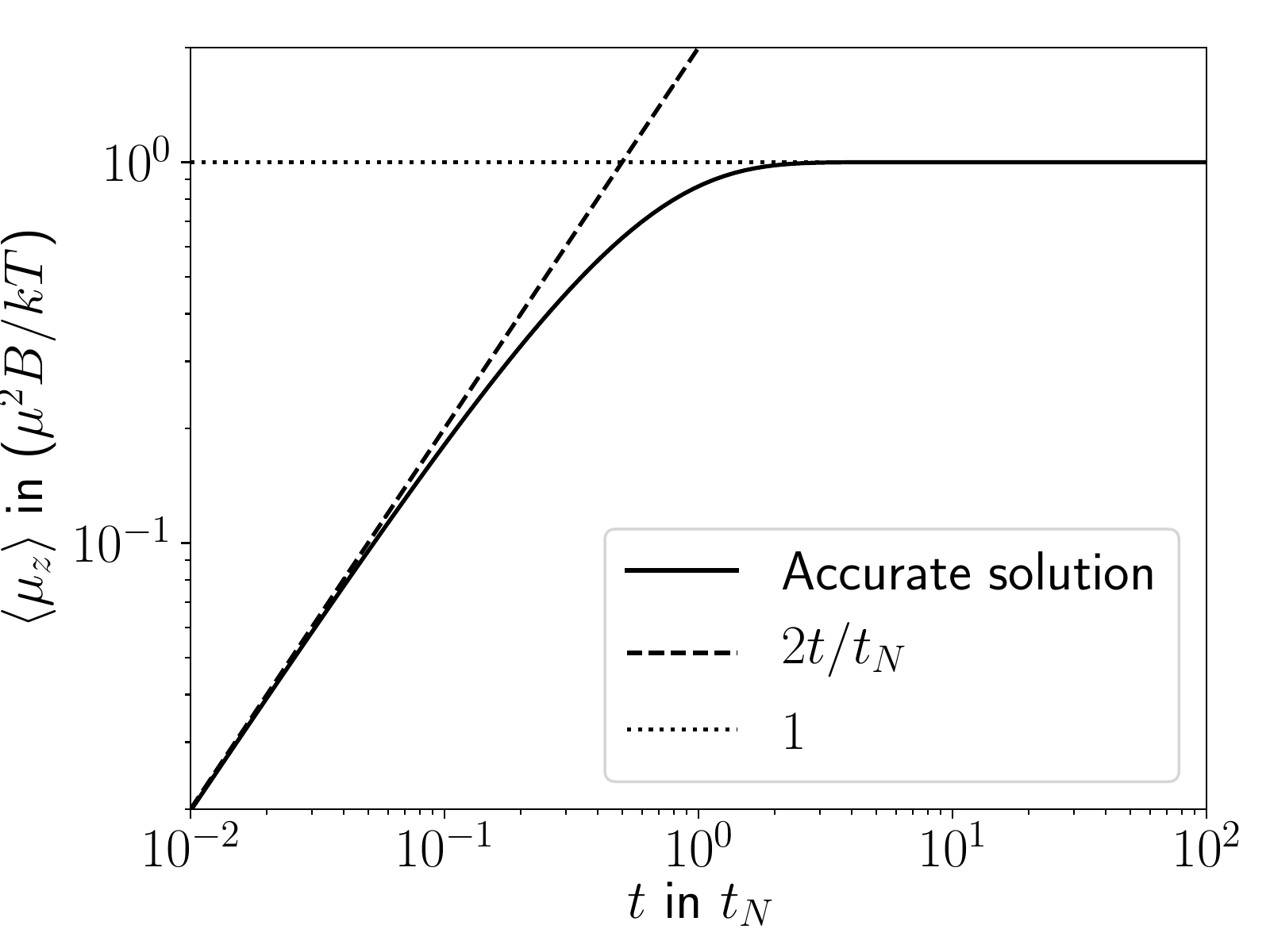}
    \caption{The averaged magnetic moment along magnetic field direction $\mean{\mu_z}$ 
    of SPIs with magnetic moment $\mu$ and N\'{e}el's relaxation timescale
    $t_N$, as a function of time $t$ exposed in an external magnetic field $B$. The SPIs considered
    here all have their easy axis along the magnetic field direction $\pm z$.}
    \label{fig:fsol}
\end{figure}

The accurate solution, together with the above two asymptotic solutions, is plotted in Fig.~\ref{fig:fsol}.
$\mu_z$ linearly increases to its equilibrium value. However, since the N\'{e}el's
relaxation timescale increases exponentially with the volume $V$, the magnetic susceptibility of
the ensemble of SPIs changes with volume $V$ very sharply across the critical volume defined in
Eq.~\eqref{eq:vcrit} which, to a good approximation, can be summarized as follows:
\begin{equation}
\chi = \left\{
\begin{array}{ll}
0, & V > V_\mathrm{cr} \\
\chi_{sp}, & V < V_\mathrm{cr}
\end{array}\right..
\end{equation}


\begin{thebibliography}{}
\expandafter\ifx\csname natexlab\endcsname\relax\def\natexlab#1{#1}\fi
\providecommand{\url}[1]{\href{#1}{#1}}
\providecommand{\dodoi}[1]{doi:~\href{http://doi.org/#1}{\nolinkurl{#1}}}
\providecommand{\doeprint}[1]{\href{http://ascl.net/#1}{\nolinkurl{http://ascl.net/#1}}}
\providecommand{\doarXiv}[1]{\href{https://arxiv.org/abs/#1}{\nolinkurl{https://arxiv.org/abs/#1}}}

\bibitem[{Abe {et~al.}(1976)Abe, Miyamoto, \& Chikazumi}]{Abe1976}
Abe, K., Miyamoto, Y., \& Chikazumi, S. 1976, Journal of the Physical Society
  of Japan, 41, 1894, \dodoi{10.1143/JPSJ.41.1894}

\bibitem[{{Alves} {et~al.}(2018){Alves}, {Girart}, {Padovani}, {Galli},
  {Franco}, {Caselli}, {Vlemmings}, {Zhang}, \& {Wiesemeyer}}]{Alves2018}
{Alves}, F.~O., {Girart}, J.~M., {Padovani}, M., {et~al.} 2018, \aap, 616, A56,
  \dodoi{10.1051/0004-6361/201832935}

\bibitem[{{Andersson} {et~al.}(2015){Andersson}, {Lazarian}, \&
  {Vaillancourt}}]{Andersson2015}
{Andersson}, B.~G., {Lazarian}, A., \& {Vaillancourt}, J.~E. 2015, \araa, 53,
  501, \dodoi{10.1146/annurev-astro-082214-122414}

\bibitem[{{Bacciotti} {et~al.}(2018){Bacciotti}, {Girart}, {Padovani}, {Podio},
  {Paladino}, {Testi}, {Bianchi}, {Galli}, {Codella}, {Coffey}, {Favre}, \&
  {Fedele}}]{Bacciotti2018}
{Bacciotti}, F., {Girart}, J.~M., {Padovani}, M., {et~al.} 2018, \apjl, 865,
  L12, \dodoi{10.3847/2041-8213/aadf87}

\bibitem[{{Bai}(2011)}]{Bai2011}
{Bai}, X.-N. 2011, \apj, 739, 50, \dodoi{10.1088/0004-637X/739/1/50}

\bibitem[{{Barnett}(1915)}]{Barnett1915}
{Barnett}, S.~J. 1915, Physical Review, 6, 239, \dodoi{10.1103/PhysRev.6.239}

\bibitem[{{Bean} \& {Livingston}(1959)}]{Bean1959}
{Bean}, C.~P., \& {Livingston}, J.~D. 1959, Journal of Applied Physics, 30,
  S120, \dodoi{10.1063/1.2185850}

\bibitem[{Billas {et~al.}(1994)Billas, Ch{\^a}telain, \& de~Heer}]{Billas1994}
Billas, I.~M., Ch{\^a}telain, A., \& de~Heer, W.~A. 1994, Science, 265, 1682,
  \dodoi{10.1126/science.265.5179.1682}

\bibitem[{{Chiang} \& {Goldreich}(1997)}]{CG1997}
{Chiang}, E.~I., \& {Goldreich}, P. 1997, \apj, 490, 368,
  \dodoi{10.1086/304869}

\bibitem[{{Coey} \& {Khalafalla}(1972)}]{Coey1972}
{Coey}, J. M.~D., \& {Khalafalla}, D. 1972, Phys. Stat. Sol. (a), 11, 229

\bibitem[{Cornell \& Schwertmann(2003)}]{IronOxides}
Cornell, R.~M., \& Schwertmann, U. 2003, The Iron Oxides: Structure,
  Properties, Reactions, Occurences and Uses (Wiley‐VCH Verlag GmbH \& Co.
  KGaA)

\bibitem[{{Cox} {et~al.}(2018){Cox}, {Harris}, {Looney}, {Li}, {Yang}, {Tobin},
  \& {Stephens}}]{Cox2018}
{Cox}, E.~G., {Harris}, R.~J., {Looney}, L.~W., {et~al.} 2018, \apj, 855, 92,
  \dodoi{10.3847/1538-4357/aaacd2}

\bibitem[{{Dai} \& {Qian}(2017)}]{Dai2017}
{Dai}, D., \& {Qian}, K. 2017, Ferromagnetism (in Chinese) (Beijing: Science
  Press)

\bibitem[{{Davis} \& {Greenstein}(1951)}]{DG1951}
{Davis}, Jr., L., \& {Greenstein}, J.~L. 1951, \apj, 114, 206,
  \dodoi{10.1086/145464}

\bibitem[{{Dent} {et~al.}(2019){Dent}, {Pinte}, {Cortes}, {M{\'e}nard},
  {Hales}, {Fomalont}, \& {de Gregorio-Monsalvo}}]{Dent2019}
{Dent}, W.~R.~F., {Pinte}, C., {Cortes}, P.~C., {et~al.} 2019, \mnras, 482,
  L29, \dodoi{10.1093/mnrasl/sly181}

\bibitem[{{Dolginov} \& {Mytrophanov}(1976)}]{DM1976}
{Dolginov}, A.~Z., \& {Mytrophanov}, I.~G. 1976, \apss, 43, 257,
  \dodoi{10.1007/BF00640009}

\bibitem[{{Draine}(1996)}]{Draine1996}
{Draine}, B.~T. 1996, Astronomical Society of the Pacific Conference Series,
  Vol.~97, {Optical and Magnetic Properties of Dust Grains}, ed. W.~G.
  {Roberge} \& D.~C.~B. {Whittet}, 16

\bibitem[{Draine(2004)}]{Draine_ColdUniverse}
Draine, B.~T. 2004, Astrophysics of Dust in Cold Clouds, ed. D.~Pfenniger \&
  Y.~Revaz (Berlin, Heidelberg: Springer Berlin Heidelberg), 213--304.
\newblock \url{https://doi.org/10.1007/3-540-31636-1_3}

\bibitem[{{Draine} \& {Hensley}(2013)}]{DraineHensley2013}
{Draine}, B.~T., \& {Hensley}, B. 2013, \apj, 765, 159,
  \dodoi{10.1088/0004-637X/765/2/159}

\bibitem[{{Draine} \& {Weingartner}(1997)}]{DW1997}
{Draine}, B.~T., \& {Weingartner}, J.~C. 1997, \apj, 480, 633,
  \dodoi{10.1086/304008}

\bibitem[{{Dutrey} {et~al.}(2014){Dutrey}, {Semenov}, {Chapillon}, {Gorti},
  {Guilloteau}, {Hersant}, {Hogerheijde}, {Hughes}, {Meeus}, {Nomura},
  {Pi{\'e}tu}, {Qi}, \& {Wakelam}}]{Dutrey2014}
{Dutrey}, A., {Semenov}, D., {Chapillon}, E., {et~al.} 2014, in Protostars and
  Planets VI, ed. H.~{Beuther}, R.~S. {Klessen}, C.~P. {Dullemond}, \&
  T.~{Henning}, 317

\bibitem[{{Goodman} \& {Whittet}(1995)}]{Goodman1995}
{Goodman}, A.~A., \& {Whittet}, D.~C.~B. 1995, \apjl, 455, L181,
  \dodoi{10.1086/309840}

\bibitem[{{Hiltner}(1949)}]{Hiltner1949nature}
{Hiltner}, W.~A. 1949, \nat, 163, 283, \dodoi{10.1038/163283a0}

\bibitem[{{Hoang}(2017)}]{Hoang2017}
{Hoang}, T. 2017, arXiv e-prints, arXiv:1704.01721.
\newblock \doarXiv{1704.01721}

\bibitem[{{Hoang} {et~al.}(2018){Hoang}, {Cho}, \& {Lazarian}}]{Hoang2018}
{Hoang}, T., {Cho}, J., \& {Lazarian}, A. 2018, \apj, 852, 129,
  \dodoi{10.3847/1538-4357/aa9edc}

\bibitem[{{Hull} \& {Zhang}(2019)}]{Hull2019}
{Hull}, C. L.~H., \& {Zhang}, Q. 2019, Frontiers in Astronomy and Space
  Sciences, 6, 3, \dodoi{10.3389/fspas.2019.00003}

\bibitem[{{Hull} {et~al.}(2018){Hull}, {Yang}, {Li}, {Kataoka}, {Stephens},
  {Andrews}, {Bai}, {Cleeves}, {Hughes}, {Looney}, {P{\'e}rez}, \&
  {Wilner}}]{Hull2018}
{Hull}, C.~L.~H., {Yang}, H., {Li}, Z.-Y., {et~al.} 2018, \apj, 860, 82,
  \dodoi{10.3847/1538-4357/aabfeb}

\bibitem[{{Iizumi} {et~al.}(1982){Iizumi}, {Koetzle}, {Shirane}, {Chikazumi},
  {Matsui}, \& {Todo}}]{Iizumi1982}
{Iizumi}, M., {Koetzle}, T.~F., {Shirane}, G., {et~al.} 1982, Acta Cryst., B38,
  2121

\bibitem[{{Jones} \& {Spitzer}(1967)}]{JS1967}
{Jones}, R.~V., \& {Spitzer}, Lyman, J. 1967, \apj, 147, 943,
  \dodoi{10.1086/149086}

\bibitem[{{Kataoka} {et~al.}(2017){Kataoka}, {Tsukagoshi}, {Pohl}, {Muto},
  {Nagai}, {Stephens}, {Tomisaka}, \& {Momose}}]{Kataoka2017}
{Kataoka}, A., {Tsukagoshi}, T., {Pohl}, A., {et~al.} 2017, \apjl, 844, L5,
  \dodoi{10.3847/2041-8213/aa7e33}

\bibitem[{Kataoka {et~al.}(2015)Kataoka, Muto, Momose, Tsukagoshi, Fukagawa,
  Shibai, Hanawa, Murakawa, \& Dullemond}]{Kataoka2015}
Kataoka, A., Muto, T., Momose, M., {et~al.} 2015, \apj, 809, 78

\bibitem[{Kataoka {et~al.}(2016)Kataoka, Tsukagoshi, Momose, Nagai, Muto,
  Dullemond, Pohl, Fukagawa, Shibai, Hanawa, \& Murakawa}]{Kataoka2016b}
Kataoka, A., Tsukagoshi, T., Momose, M., {et~al.} 2016, \apj, 831, L12

\bibitem[{{Lazarian}(2007)}]{Lazarian2007}
{Lazarian}, A. 2007, \jqsrt, 106, 225, \dodoi{10.1016/j.jqsrt.2007.01.038}

\bibitem[{{Lazarian} \& {Hoang}(2007)}]{LH2007}
{Lazarian}, A., \& {Hoang}, T. 2007, \mnras, 378, 910,
  \dodoi{10.1111/j.1365-2966.2007.11817.x}

\bibitem[{{Li} {et~al.}(2016){Li}, {Pantin}, {Telesco}, {Zhang}, {Wright},
  {Barnes}, {Packham}, \& {Mari{\~n}as}}]{Li2016}
{Li}, D., {Pantin}, E., {Telesco}, C.~M., {et~al.} 2016, \apj, 832, 18,
  \dodoi{10.3847/0004-637X/832/1/18}

\bibitem[{Li {et~al.}(2007)Li, Montano, Barbiellini, Mijnarends, Kaprzyk, \&
  Bansil}]{LI20071556}
Li, Y., Montano, P., Barbiellini, B., {et~al.} 2007, Journal of Physics and
  Chemistry of Solids, 68, 1556 ,
  \dodoi{https://doi.org/10.1016/j.jpcs.2007.03.037}

\bibitem[{{Lin} {et~al.}(2019){Lin}, {Li}, {Yang}, {Looney}, {Stephens}, \&
  {Hull}}]{Lin2019}
{Lin}, Z.-Y.~D., {Li}, Z.-Y., {Yang}, H., {et~al.} 2019, arXiv e-prints,
  arXiv:1912.10012.
\newblock \doarXiv{1912.10012}

\bibitem[{{Mathewson} \& {Ford}(1970)}]{Mathewson1970}
{Mathewson}, D.~S., \& {Ford}, V.~L. 1970, \memras, 74, 139

\bibitem[{{Mathis}(1986)}]{Mathis1986}
{Mathis}, J.~S. 1986, \apj, 308, 281, \dodoi{10.1086/164499}

\bibitem[{Morrish(2001)}]{Morrish2001}
Morrish, A.~H. 2001, The Physical Principles of Magnetism (Wiley-IEEE Press)

\bibitem[{{N\'{e}el}(1949)}]{Neel1949}
{N\'{e}el}, L. 1949, Ann. G\'{e}ophys., 5, 99

\bibitem[{{Ohashi} \& {Kataoka}(2019)}]{Ohashi2019}
{Ohashi}, S., \& {Kataoka}, A. 2019, \apj, 886, 103,
  \dodoi{10.3847/1538-4357/ab5107}

\bibitem[{{Ohashi} {et~al.}(2018){Ohashi}, {Kataoka}, {Nagai}, {Momose},
  {Muto}, {Hanawa}, {Fukagawa}, {Tsukagoshi}, {Murakawa}, \&
  {Shibai}}]{Ohashi2018}
{Ohashi}, S., {Kataoka}, A., {Nagai}, H., {et~al.} 2018, \apj, 864, 81,
  \dodoi{10.3847/1538-4357/aad632}

\bibitem[{Pisane {et~al.}(2017)Pisane, Singh, \& Seehra}]{Pisane2017}
Pisane, K.~L., Singh, S., \& Seehra, M.~S. 2017, Applied Physics Letters, 110,
  222409, \dodoi{10.1063/1.4984903}

\bibitem[{{Planck Collaboration} {et~al.}(2015){Planck Collaboration}, {Ade},
  {Aghanim}, {Alina}, {Alves}, {Armitage-Caplan}, {Arnaud}, {Arzoumanian},
  {Ashdown}, {Atrio-Barand ela}, {Aumont}, {Baccigalupi}, {Banday}, {Barreiro},
  {Battaner}, {Benabed}, {Benoit-L{\'e}vy}, {Bernard}, {Bersanelli},
  {Bielewicz}, {Bock}, {Bond}, {Borrill}, {Bouchet}, {Boulanger}, {Bracco},
  {Burigana}, {Butler}, {Cardoso}, {Catalano}, {Chamballu}, {Chary}, {Chiang},
  {Christensen}, {Colombi}, {Colombo}, {Combet}, {Couchot}, {Coulais}, {Crill},
  {Curto}, {Cuttaia}, {Danese}, {Davies}, {Davis}, {de Bernardis}, {de Gouveia
  Dal Pino}, {de Rosa}, {de Zotti}, {Delabrouille}, {D{\'e}sert}, {Dickinson},
  {Diego}, {Donzelli}, {Dor{\'e}}, {Douspis}, {Dunkley}, {Dupac}, {Efstathiou},
  {En{\ss}lin}, {Eriksen}, {Falgarone}, {Ferri{\`e}re}, {Finelli}, {Forni},
  {Frailis}, {Fraisse}, {Franceschi}, {Galeotta}, {Ganga}, {Ghosh}, {Giard},
  {Giraud-H{\'e}raud}, {Gonz{\'a}lez-Nuevo}, {G{\'o}rski}, {Gregorio},
  {Gruppuso}, {Guillet}, {Hansen}, {Harrison}, {Helou},
  {Hern{\'a}ndez-Monteagudo}, {Hildebrand t}, {Hivon}, {Hobson}, {Holmes},
  {Hornstrup}, {Huffenberger}, {Jaffe}, {Jaffe}, {Jones}, {Juvela},
  {Keih{\"a}nen}, {Keskitalo}, {Kisner}, {Kneissl}, {Knoche}, {Kunz},
  {Kurki-Suonio}, {Lagache}, {L{\"a}hteenm{\"a}ki}, {Lamarre}, {Lasenby},
  {Lawrence}, {Leahy}, {Leonardi}, {Levrier}, {Liguori}, {Lilje},
  {Linden-V{\o}rnle}, {L{\'o}pez-Caniego}, {Lubin}, {Mac{\'\i}as-P{\'e}rez},
  {Maffei}, {Magalh{\~a}es}, {Maino}, {Mandolesi}, {Maris}, {Marshall},
  {Martin}, {Mart{\'\i}nez-Gonz{\'a}lez}, {Masi}, {Matarrese}, {Mazzotta},
  {Melchiorri}, {Mendes}, {Mennella}, {Migliaccio}, {Miville-Desch{\^e}nes},
  {Moneti}, {Montier}, {Morgante}, {Mortlock}, {Munshi}, {Murphy}, {Naselsky},
  {Nati}, {Natoli}, {Netterfield}, {Noviello}, {Novikov}, {Novikov},
  {Oxborrow}, {Pagano}, {Pajot}, {Paladini}, {Paoletti}, {Pasian}, {Pearson},
  {Perdereau}, {Perotto}, {Perrotta}, {Piacentini}, {Piat}, {Pietrobon},
  {Plaszczynski}, {Poidevin}, {Pointecouteau}, {Polenta}, {Popa}, {Pratt},
  {Prunet}, {Puget}, {Rachen}, {Reach}, {Rebolo}, {Reinecke}, {Remazeilles},
  {Renault}, {Ricciardi}, {Riller}, {Ristorcelli}, {Rocha}, {Rosset},
  {Roudier}, {Rubi{\~n}o-Mart{\'\i}n}, {Rusholme}, {Sandri}, {Savini}, {Scott},
  {Spencer}, {Stolyarov}, {Stompor}, {Sudiwala}, {Sutton}, {Suur-Uski},
  {Sygnet}, {Tauber}, {Terenzi}, {Toffolatti}, {Tomasi}, {Tristram}, {Tucci},
  {Umana}, {Valenziano}, {Valiviita}, {Van Tent}, {Vielva}, {Villa}, {Wade},
  {Wandelt}, {Zacchei}, \& {Zonca}}]{Planck2015XIX}
{Planck Collaboration}, {Ade}, P.~A.~R., {Aghanim}, N., {et~al.} 2015, \aap,
  576, A104, \dodoi{10.1051/0004-6361/201424082}

\bibitem[{{Purcell}(1979)}]{Purcell1979}
{Purcell}, E.~M. 1979, \apj, 231, 404, \dodoi{10.1086/157204}

\bibitem[{{Roberge} {et~al.}(1993){Roberge}, {Degraff}, \&
  {Flaherty}}]{Roberge1993}
{Roberge}, W.~G., {Degraff}, T.~A., \& {Flaherty}, J.~E. 1993, \apj, 418, 287,
  \dodoi{10.1086/173390}

\bibitem[{Stephens {et~al.}(2014)Stephens, Looney, Kwon,
  Fern{\'{a}}ndez-L{\'{o}}pez, Hughes, Mundy, Crutcher, Li, \&
  Rao}]{Stephens2014}
Stephens, I.~W., Looney, L.~W., Kwon, W., {et~al.} 2014, \nat, 514, 597

\bibitem[{{Stephens} {et~al.}(2017){Stephens}, {Yang}, {Li}, {Looney},
  {Kataoka}, {Kwon}, {Fern{\'a}ndez-L{\'o}pez}, {Hull}, {Hughes}, {Segura-Cox},
  {Mundy}, {Crutcher}, \& {Rao}}]{Stephens2017}
{Stephens}, I.~W., {Yang}, H., {Li}, Z.-Y., {et~al.} 2017, \apj, 851, 55,
  \dodoi{10.3847/1538-4357/aa998b}

\bibitem[{{Tazaki} {et~al.}(2017){Tazaki}, {Lazarian}, \&
  {Nomura}}]{Tazaki2017}
{Tazaki}, R., {Lazarian}, A., \& {Nomura}, H. 2017, \apj, 839, 56,
  \dodoi{10.3847/1538-4357/839/1/56}

\bibitem[{{Tobin} {et~al.}(2020){Tobin}, {Sheehan}, {Megeath},
  {D{\'\i}az-Rodr{\'\i}guez}, {Offner}, {Murillo}, {van 't Hoff}, {van
  Dishoeck}, {Osorio}, {Anglada}, {Furlan}, {Stutz}, {Reynolds}, {Karnath},
  {Fischer}, {Persson}, {Looney}, {Li}, {Stephens}, {Chand ler}, {Cox},
  {Dunham}, {Tychoniec}, {Kama}, {Kratter}, {Kounkel}, {Mazur}, {Maud},
  {Patel}, {Perez}, {Sadavoy}, {Segura-Cox}, {Sharma}, {Stephenson}, {Watson},
  \& {Wyrowski}}]{Tobin2020}
{Tobin}, J.~J., {Sheehan}, P.~D., {Megeath}, S.~T., {et~al.} 2020, \apj, 890,
  130, \dodoi{10.3847/1538-4357/ab6f64}

\bibitem[{{Valdivia} {et~al.}(2019){Valdivia}, {Maury}, {Brauer}, {Hennebelle},
  {Galametz}, {Guillet}, \& {Reissl}}]{Valdivia2019}
{Valdivia}, V., {Maury}, A., {Brauer}, R., {et~al.} 2019, \mnras, 488, 4897,
  \dodoi{10.1093/mnras/stz2056}

\bibitem[{{Verwey}(1939)}]{Verwey1939}
{Verwey}, E. J.~W. 1939, Nature, 144, 327

\bibitem[{Walz(2002)}]{Walz2002}
Walz, F. 2002, Journal of Physics: Condensed Matter, 14, R285,
  \dodoi{10.1088/0953-8984/14/12/203}

\bibitem[{{Weidenschilling}(1977)}]{Weidenschilling1977}
{Weidenschilling}, S.~J. 1977, \apss, 51, 153, \dodoi{10.1007/BF00642464}

\bibitem[{Yang {et~al.}(2016a)Yang, Li, Looney, \& Stephens}]{Yang2016a}
Yang, H., Li, Z.-Y., Looney, L., \& Stephens, I. 2016a, \mnras, 456, 2794

\bibitem[{Yen {et~al.}(2017)Yen, Koch, Takakuwa, Krasnopolsky, Ohashi, \&
  Aso}]{Yen2017}
Yen, H.-W., Koch, P.~M., Takakuwa, S., {et~al.} 2017, The Astrophysical
  Journal, 834, 178, \dodoi{10.3847/1538-4357/834/2/178}

\end{thebibliography}
\end{document}